# ADVANCED INFLATABLE DE-ORBIT SOLUTIONS FOR DERELICT SATELLITES AND ORBITAL DEBRIS

## Aman Chandra,[*] Greg Wilburn,[†] and Jekan Thangavelautham[‡]

The exponential rise in small-satellites and CubeSats in Low Earth Orbit (LEO) poses important challenges for future space traffic management. At altitudes of 600 km and lower, aerodynamic drag accelerates de-orbiting of satellites. However, placement of satellites at higher altitudes required for constellations pose important challenges. The satellites will require on-board propulsion to lower their orbits to 600 km and let aerodynamic drag take-over. In this work we analyze solutions for de-orbiting satellites at altitudes of up to 3000 km. We consider a modular robotic de-orbit device that has stowed volume of a regular CubeSat. The de-orbit device would be externally directed towards a dead satellite or placed on one by an external satellite servicing system. Our solutions are intended to be simple, high-reliability devices that operate in a passive manner, requiring no active electronics or utilize external beamed power in the form of radio frequency, microwave or laser to operate. Utilizing this approach, it is possible for an external, even ground based system to direct the de-orbit of a spacecraft. The role of an external system to direct the de-orbit is important to avoid accidental collisions. Some form of propulsion is needed to lower the orbit of the dead satellite or orbital debris. We considered green (non-toxic) propulsion methods including solar radiation pressure, solar-thermal propulsion using water steam, solar-electrolysis propulsion using water and use of electrodynamic tethers. Based on this trade-study we identify multiple solutions that can be used to de-orbit a spacecraft or orbital debris.

**1.0 INTRODUCTION**

CubeSats missions in space are steadily increasing. CubeSat platforms have gained popularity due to the low-cost access to space they provide. With physical dimensions conforming to reference standards [1]. This has also in-turn greatly reduced development cost and engineering required. Furthermore, recent advances in radiation hardened micro-electronics [2] have allowed CubeSat electronics to be used reliably in Geosynchronous Earth Orbits (GEO) and beyond. The successful launch and operation of the MarCO CubeSats in 2018 is enabling evaluation of CubeSat subsystems and electronics in deep space [3].

There has been an exponential increase in the number of CubeSat missions being sent into space since 2010. Studies have shown that increasing launch opportunities have led to the placement of 60-70 new objects per year into LEO [4]. It is necessary, therefore, to provide traffic management and service Low Earth Orbits to remove orbiting debris. Fig. 1 is a visual illustration of LEO congestion gone awry.

[*] PhD Student, Aerospace and Mechanical Engineering, University of Arizona, 85721, USA.
[†] PhD Student, Aerospace and Mechanical Engineering, University of Arizona, 85721, USA.
[‡] Assistant Professor, Aerospace and Mechanical Engineering, University of Arizona, 85721, USA.

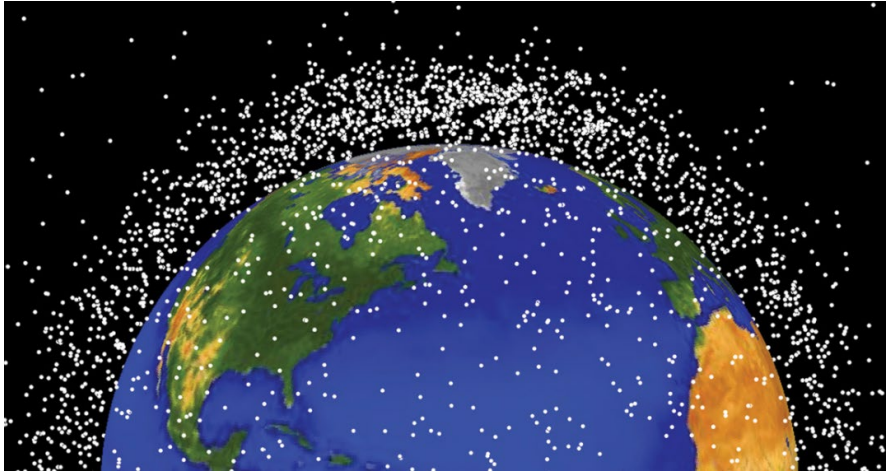

Fig. 1. Artist's impression of growing debris

The most popular strategies for orbit servicing include Active Debris Removal (ADR) and On-Orbit Satellite Servicing (OSS) [5]. Both methods faces increased mission complexity, cost and logistics. Constraints imposed by limited volume and mass available on CubeSats makes them unfeasible. A far more efficient strategy is to design an orbital system capable of de-orbiting and 'self-servicing' at the end of the mission. CubeSats in Low Earth Orbit encounter aerodynamic drag leading to orbital decay over time resulting in burn up and disposal in the Earth's atmosphere. For altitudes above 500 km, however, atmospheric density reduces significantly.

This could lead a CubeSat to remain in orbit beyond 25 years, the current mandated limit for orbit lifetimes. Increasing encountered drag for small satellites is key to reducing their orbital lifetimes. Thrusters have been used to provide propulsive deceleration. In the case of CubeSats, however, including a propulsion system and allocating a considerable volume for propellant may not always be practical. This leads us to deployable structures as an alternative to achieve required aerodynamic drag. Among deployable structures, inflatable gossamers offer the highest packing efficiency. Such structures are ultra-light and offer very high deployed surface to mass ratios. Additionally, they can be scaled to sizes in the order of meters. These attributes make it possible to achieve low ballistic coefficients with inflatables. Table 1 shows a comparison of various deployable technologies for braking.

Table 1. Comparison of deployable braking technologies

| Braking technology | Mass/ unit surface area (kg) | Packing ratio | Drag co-efficient | Ballistic co-efficient (Pa) |
|---|---|---|---|---|
| Deployable panels | 0.5 – 0.7 | 2:1 – 3:1 | 2 - 4 | 150 - 200 |
| Linkage systems | 0.7 - 1 | 5:1 – 8:1 | 3 - 4 | 20 - 100 |
| Inflatable gossamers | 0.06 – 0.2 | 15:1 – 20:1 | 2 - 4 | 5 - 20 |

In this paper, we analyze inflatable aerodynamic drag/braking devices. A design strategy is evolved to produce conceptual braking structures compatible with the CubeSat form factor. We also consider tethering technologies that can work in conjunction with inflatable decelerators to provide an efficient solution for debris capture and disposal. We compute the drag co-efficient and ballistic co-efficient of the proposed design to understand expected braking performance. In the following sections we presented related work, followed by methodology, results, discussion and conclusions/future work.

## 2.0 BACKGROUND

Inflatable technology has been the subject of investigation since the 1950's. First major success came with NASA's ECHO balloon project [1]. Two inflatable Mylar balloons each spanning several meters in diameter successfully operated in the upper atmosphere over their designated mission spans. This established membrane-based inflatables as a reliable technology in space. Several structural applications have been studied since, varying from gossamer sails, antennas, landing airbags and solar panels [2]. Ruggedized inflatables made of thermal fabrics started being researched for the challenging thermo-structural conditions during atmospheric re-entry. The first inflatable re-entry test was carried out in the year 2000 as a demonstration of inflatable re-entry and descent technology (IRDT) [3]. Structural and thermal performance was observed to be enough to survive atmospheric-entry.

A number of studies have been carried out on inflatable devices to facilitate atmospheric entry. Successful tests were carried out from the inflatable re-entry vehicle experiment (IRVE) in 2006. The purpose of the experiment was to validate aero-shell performance for atmospheric re-entry [4]. While atmospheric entry has received considerable attention, the use of inflatable structures as drag devices has seen much fewer efforts. Andrews Space has developed a prototype that has undergone ground based tests as an inflatable nanosat de-orbit and recovery system for CubeSat payloads [5]. Fig. 3 shows an illustration of their design. Italian concept IRENE [6] is undergoing tests with a spherical cone designs but is intended for much larger payloads. The structural principles towards designing a drag and de-orbit device are fairly-well understood [7]. The critical parameter is the ballistic coefficient of these structures. Lower ballistic coefficients lead to a reduction in mechanical and thermal loads experienced by the structure while achieving higher deceleration rates. This highlights the potential of using large inflatable gossamers that can be packaged into very small volumes for nano-satellite payloads.

Some conceptual studies on inflatable drag devices have been conducted. Among large scale gossamer structures, Global Aerospace Corporation proposed the Gossamer Orbit Lowering Device (GOLD) to de-orbit spent stages and old or derelict satellites [8]. The concept consists of deploying a large inflatable sphere several meters in diameter that offers exceptionally low ballistic coefficients in Lower Earth Orbit. Gossamer sails made of Kapton have also been studied in considerable detail [9]. While sails potentially offer more efficient packing ratios than inflatables, their structural reliability for aerobraking is not well established. While encountering loads due to atmospheric drag, a pneumatic pressure system has been used to provide necessary resistive stiffness. In the case of sails, additional structural re-enforcement is needed which reduces packing efficiency and increases deployment complexity. Pneumatic inflatables have shown robust structural behavior while maintaining ease of scaling into large sizes. The focus of our research is on inflatable structures. Pneumatic inflatable require a gas source. This can be in the form of a compressed gas or gas producing chemical reaction. Inflatables using solid state sublimates as gas sources have shown promising results for Low Earth Orbit operations [10-12].

Tethers and tethering mechanisms have also received limited attention as a device on board a small satellite. Tethers can also be classified as gossamers due to their lightweight construction [3]. However, their structural reliability exists only in tension as they do not possess the ability to resist compressive loads.

## 3.0 CUBESAT DRAG DEVICE DESIGN AND ANALYSIS

De-orbit performance is proportional to the drag force experienced by the device. As the spacecraft altitude increases, the surrounding atmosphere continues to rarify thereby reducing drag. The effect of height on atmospheric drag is described by (1)

$$\rho \approx \rho_o e^{-\Delta h / h_o(h)} \quad (1)$$

The atmospheric density $\rho$ at a given altitude and $\rho_o$ at a second altitude with difference in height of $\Delta h$ are related exponentially as shown. $h_o(h)$ termed as scale height is a function of altitude. We begin by studying the nature of forces encountered towards two major applications. For a circular orbit at altitude $H$ above the Earth, the average change in acceleration due to drag is as shown in (2)

$$\Delta a_{rev} = -\frac{2\pi \rho a^2}{b_c} \quad (2)$$

To enable atmospheric burn up at 100 km altitude, requires the ballistic coefficient be a function of altitude [13] and can be written as:

$$b_c = \frac{(R_e + H)^2 (2R_e + H + 10^5)(H - 10^5)}{2\pi \mu \rho (R_e + 10^5)^2} \quad (3)$$

We propose designs for aero-braking structures on board CubeSats. Design elements for such systems have been described in detail [17, 18] for rigid structures. We apply similar principles to gossamer membranes. Traditionally, two separate strategies have been developed. One is the design on the support structure and the other is that of the shield that offers resistance and drag. The support truss is designed to transfer aerodynamic drag from the braking shield structure. Fig. 2 shows a schematic diagram of this basic configuration:

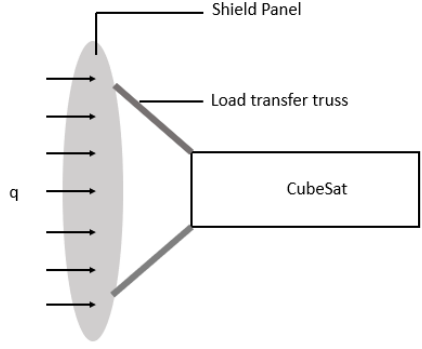

Fig. 2: Structural design configuration

The pressure distribution on the surface of the shield can be assumed to be uniform in nature and is calculated from the aerobrake inertia force due to a constant rate of deceleration. The pressure $p$ is given as shown in (4) as:

$$p = \frac{M_s a}{g A_{ab}} \qquad (4)$$

Here $M_s$ is the mass of the spacecraft attached to the braking system, $a$ is the deceleration rate, $g$ is a constant defined by Newton's law as Force = (mass × acceleration). $A_{ab}$ is the area of the aerobraking structure. We extend their methodology to include design of inflatable membrane structural units. Based upon structural function, the aero-braking device consists of a shield and support structure. The fundamental structural sizing equation is a shown below.

$$w_{max} = \alpha \frac{qA^4}{D_{HP}} \qquad (5)$$

Here $w_{max}$ represents a bound on maximum mass of the structure for achieving a bending stiffness $D_{HP}$ for encountered drag force $q$ over area $A$. Based on sizing requirements, we propose two spherical cone based structural concepts [13] as shown in Fig. 3 and 4 below:

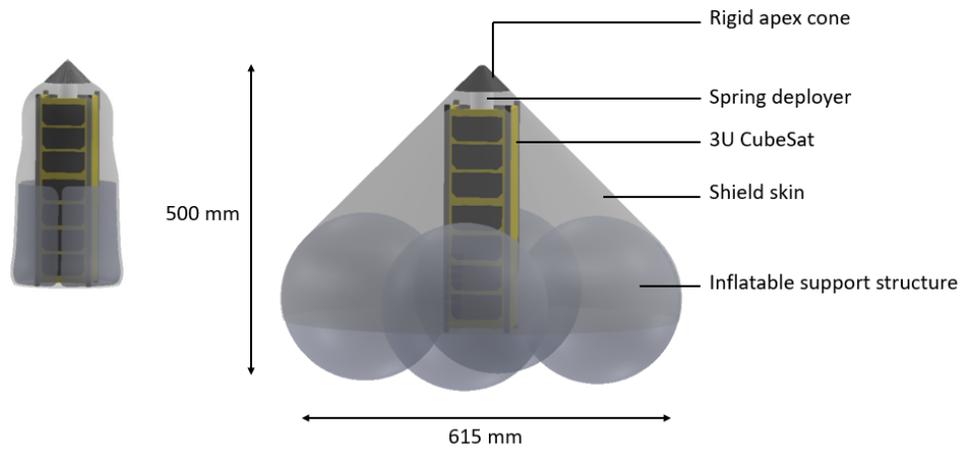

Fig. 3. Concept 1 with spherical inflatable aeroshell

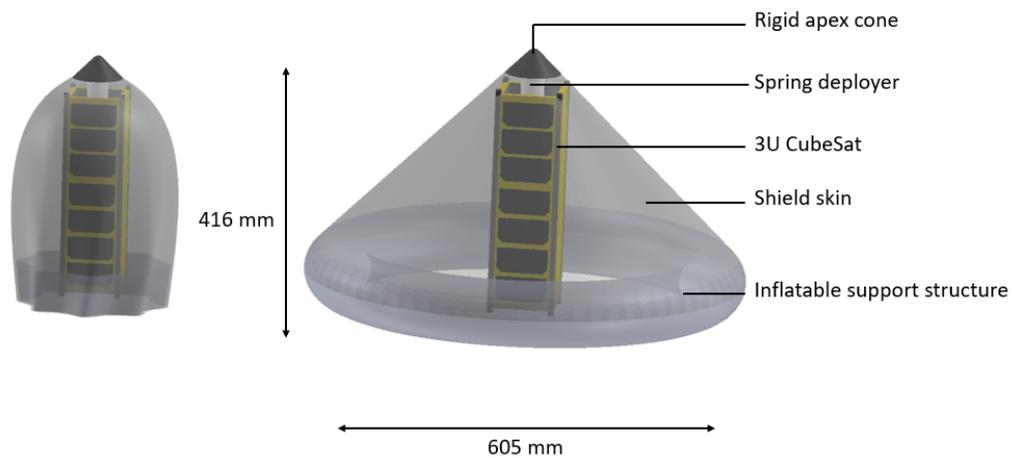

Fig. 4. Concept 2 with toroidal inflatable aero-shell

Our concepts consist of inflatable structural members providing load transfer from a membrane skin. The apex cone acts as rigid support on the other end. The rigidity of the apex is to withstand thermo-structural load concentration at the apex region. Concept 1 uses an assembly of inflatable spheres as support while in Concept 2 that has been replaced by an inflatable toroid.

Aero-braking or de-orbit performance of the inflatable is characterized by estimating its drag co-efficient. The following equation is used to assess the drag co-efficient based on the structure's geometry:

$$C_D = \alpha \left( \frac{4}{A} \int_S \cos^3 \phi \, dS \right) + \beta \left( 2 + \frac{1}{A} \int_S \cos^2 \phi \, dS \right) + 2\gamma \qquad (6)$$

*α, β* and *γ* represent relative fractions contributing to specular, diffuse and absorptive surfaces on the structure. Hence, the sum *α+β+γ* equals 1. *Φ* represents the angle between the structure's surface normal and velocity vector. Computed drag coefficient values were used to calculated obtained deceleration using (7).

$$\Delta P_{rev} = -6\pi^2 \left( C_D A / m \right) \rho a^2 / V \qquad (7)$$

Here *ΔP* represents a change in orbital period for a circular orbit characterized by the CubeSats velocity V and mass m. The calculated loads are compared with expected structural behavior to understand their ability to maintain structural integrity. Table 2 shows estimated drag coefficients, ballistic coefficients and estimated orbit decay for a 3U CubeSat using both inflatable concepts with an estimated total mass of 4 kg.

Table 2. Comparison of braking concepts

| Design Concept | Mass (kg) | Surface Area (m²) | Drag Co-efficient | Ballistic Co-efficient (kg/m²) |
|---|---|---|---|---|
| 3U CubeSat | 3.5 | 0.01 | 2 | 175 |
| Concept 1 | 4 | 0.248 | 2.667 | 6.05 |
| Concept 2 | 4 | 0.346 | 2.7 | 4.28 |

The above table shows a dramatic decrease in ballistic co-efficient upon adding the inflatable structures onto the 3U CubeSat. This is due to much larger surface areas at very low additional mass. A larger drag coefficient is possible in thanks to an optimized spherical cone geometry. Based on calculated co-efficient values, we go on to calculate estimated de-orbit lifetimes for each case. Table 3 shows estimated orbit decay times from various altitudes of a circular orbit for both design concepts.

Table 3. Comparison of expected de-orbit times

| Altitude (km) | Disposal Life-time (years) | | |
|---|---|---|---|
| | 3U CubeSat | Concept 1 | Concept 2 |
| 400 | 1.2 | 0.05 | 0.045 |
| 500 | 6.3 | 0.3 | 0.28 |
| 600 | 23.5 | 1.58 | 1.5 |
| 700 | >25 | 6.4 | 6.2 |
| 800 | >25 | 18.5 | 17.8 |

**4.0 INTEGRATED TETHER AND DRAG DEVICE CONCEPT**

Inflatables are excellent as de-orbit devices but not for capture of space-debris. This is due to their limited structural stiffness and susceptibility to puncture and damage. Tethers on the other hand can be constructed of highly toughened tensile fibers that have the ability to absorb the shock associated with the capture of large derelict objects. We propose an integrated CubeSat system consisting of a tether to facilitate debris capture and a drag device to facilitate debris disposal. Figure 5 shows a model of the proposed system.

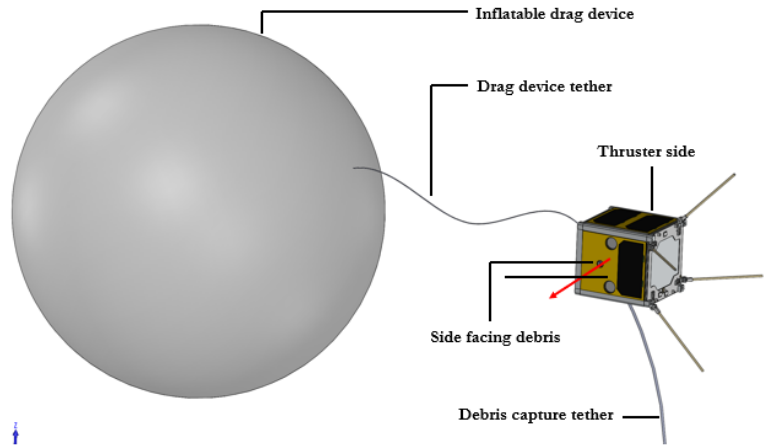

Fig. 5. Concept 2 with toroidal inflatable aero-shell

Preliminary studies have shown that a Vectran fiber tether of approximately 2 meters length can be packaged into 0.2U of volume. Further, a drag device in the form of an inflated sphere of diameter 0.5m can be package into 0.3U volume. Based on these estimates, it is possible to package both these systems on board a 1U CubeSat bus. This configuration has been chosen as the baseline for our analysis.

**5.0 CONCLUSIONS**

Our work demonstrates design concepts for inflatable membrane drag devices on board small satellite and CubeSat platforms. First order estimates of expected drag enhancements show potential for this technology as a low-cost solution to on-orbit servicing. A drag device attached to potentially derelict spacecraft can vastly simplify the process of debris disposal. We also extend the concept to include an integrated tether system to capture debris of irregular geometry. Preliminary analysis shows the feasibility of such a system. Future work includes incorporating high-fidelity atmospheric models to estimate drag coefficients with greater accuracy. This would be used to further refine the structural design of the proposed concepts. The structural model will need modifications to incorporate thermal loads and thermal stress concentrations on the inflatable device. This will be followed by testing in a laboratory followed by testing under a relevant environment.